\documentclass[mathleft
]{an}
\usepackage{graphicx}
\usepackage{times}
\begin{document}

\Pagespan{1}{}
\Yearpublication{...}%
\Yearsubmission{...}%
\Month{...}%
\Volume{...}%
\Issue{...}%

\title{Relaxation of protostellar accretion shocks using the smoothed particle hydrodynamics}

\author{M. Nejad-Asghar\thanks{ \email{nejadasghar@umz.ac.ir}\newline}
}
\titlerunning{Relaxation of accretion shocks}
\authorrunning{M. Nejad-Asghar}
\institute{ Department of Physics, University of Mazandaran,
Babolsar, Iran}

\received{...} \accepted{...} \publonline{...}


\abstract{%
It is believed that protostellar accretion disks to be formed from
nearly ballistic infall of the molecular matters in rotating core
collapse. Collisions of these infalling matters lead to formation of
strong supersonic shocks, which if they cool rapidly, result in
accumulation of that material in a thin structure in the equatorial
plane. Here, we investigate the relaxation time of the protostellar
accretion post-shock gas using the smoothed particle hydrodynamics
(SPH). For this purpose, a one-dimensional head-on collision of two
molecular sheets is considered, and the time evolution of the
temperature and density of the post-shock region simulated. The
results show that in strong supersonic shocks, the temperature of
the post-shock gas quickly increases proportional to square of the
Mach number, and then gradually decreases according to the cooling
processes. Using a suitable cooling function shows that in
appropriate time-scale, the center of the collision, which is at the
equatorial plane of the core, is converted to a thin dense molecular
disk, together with atomic and ionized gases around it. This
structure for accretion disks may justify the suitable conditions
for grain growth and formation of proto-planetary entities.}

\maketitle

\section{Introduction}
In the general picture, the gravitational unstable molecular cloud
cores collapse and form the protostars, accretion disks, and
outflows. The first model of a spherical symmetric accretion which
flow towards a central object was made by Bondi~(1952). Years later,
Ulrich~(1976) modified the Bondi's idea by assuming all fluid
particles have a certain angular momentum with negligible pressure
forces at the border of the cloud, so that the collapse problem
analyzed using ballistic trajectories. Since then, many
generalizations of the core collapse have been made (e.g., Shu~1977,
Terebey, Shu \& Cassen~1984, Foster \& Chevalier~1993, Galli \&
Shu~1993, Henriksen, Andr\'{e} \& Bontemps~1997, Fatuzzo, Adams \&
Myers~2004, Mendoza, Tejeda \& Nagel~2009, Bate~2010,
Nejad-Asghar~2010). As a result of rotating collapse, the collisions
at the equatorial plane, between the infalling matters coming from
the northern hemisphere and those coming from the southern one,
cause the formation of strong supersonic shocks. If the shocked gas
cools rapidly, the result is that material accumulates in a thin
structure in the equatorial plane, i.e., accretion disk
(Hartmann~2009).

Observations have revealed not only the existence of the
circumstellar disks around the protostars (e.g., Watson et al.~2007,
Akeson~2008, Quanz et al.~2010), but also the occurrence of jets and
outflows (e.g., Hirth, Mundt \& Solf 1997, Wu et al.~2004, Dunham et
al.~2010). The outflows and jets are ubiquitous together with
accretion disks through the collapsing molecular cloud cores (e.g.,
Furuya, Cesaroni \& Shinnaga~2011). We cannot directly observe the
newborn or very young accretion disks and jets because their
formation places are embedded in a dense infalling envelope. On the
other words, observations cannot directly determine the real
structure and formation history of accretion disks and outflows.
Therefore, theoretical approach and simulations are necessary to
investigate the formation and evolution of them (e.g., Walch et
al.~2009, Machida, Inutsuka \& Matsumoto~2010, Ciardi \&
Hennebelle~2010, Nejad-Asghar~2011). Clearly, consideration of jets
and magnetic fields can affect on formation and evolution of
accretion disks (e.g., Hujeirat~1998, Hujeirat et al.~2003, Machida,
Inutsuka \& Matsumoto~2007), but in this study, we turn our
attention to the phase after formation of of central protostar, and
for simplicity neglect the effect of jets and outflows on
protostellar accretion shocks. In addition to physical evolution of
circumstellar disks, the chemical evolution from the core to the
disk phase is also important (e.g., Ceccarelli, Hollenbach \&
Tielens~1996, Rodgers \& Charnley~2003, Garrod \& Herbst~2006,
Garrod, Weaver \& Herbst 2008, Visser et al.~2009). Here, we assume
that all effects of chemical evolution of shocked gas are simplified
through the appropriate net cooling function.

Interstellar shocks cover a wide range of parameters: velocities of
$1-10^4 \mathrm{km.s^{-1}}$, pre-shock densities of $10^{-2}-10^7
\mathrm{cm^{-3}}$, and post-shock temperatures of $10^2-10^9
\mathrm{K}$. The strength of a shock is indicated by the Mach number
which can range up to $\sim 10^3$, for larger than laboratory shocks
(McKee \& Hollenbach~1980). For an adiabatic strong shock, the jump
in density is limited to a factor of four (for a ratio of specific
heats $\gamma=5/3$), while the temperature of post-shock gas can be
increased proportional to the square of Mach number (e.g., Dyson \&
Williams~1997). As mentioned above, the high velocity of infalling
matters at the equatorial plane of a rotating core collapse, leads
to strong supersonic collisions with great Mach numbers. The high
temperature of the post-shock gas causes the molecule dissociation
and the atom ionization. Neufeld \& Hollenbach (1994) examined the
physical and chemical processes of high-density accretion shocks
which associated with the supersonic infall of material during the
collapse of a molecular cloud core to form a protostar. Since the
rotation and magnetic fields cause the deflection of infalling
matters from the protostar so that a shocked disk gas is formed,
here we study the physical and chemical evolutions of these shocked
circumstellar disks for understanding the formation of structure and
substructures through them.

As a general aspect of star formation, we expect that cooling (as a
result of chemical and physical changes) of the protostellar
accretion shocked gas at the equatorial plane of collapsing core
leads to supply the suitable conditions for grain growth and
formation of proto-planetary entities. The goal of this paper is to
investigate this expectation. For this purpose, the jump shock
(J-shock) structure and the cooling time-scale of post-shock gas are
presented in section~2. In section~3, we use the SPH methodology to
investigate the time evolution of the strong supersonic shocks.
Finally, section~4 is devoted to summary and conclusions.

\section{J-shock structure}

For simplicity, the shock is assumed planar and steady in which the
deceleration is negligible and there is no thermal instability in
the cooling layer. The jump conditions of this shock (J-shock)
relate the quantities at an arbitrary point behind the shock front
to those ahead of it. Conservation of mass, momentum, and energy
across the shock front is given by Rankine-Hugoniot conditions
(e.g., Dyson \& Williams~1997)
\begin{equation}\label{rh1}
\rho_1 v_1=\rho_2 v_2
\end{equation}
\begin{equation}\label{rh2}
\rho_1 v_1^2+ \mathcal{K}_1 \rho_1 T_1 =\rho_2 v_2^2+ \mathcal{K}_2
\rho_2 T_2
\end{equation}
\begin{equation}\label{rh3}
\frac{1}{2}v_1^2 + \frac{\gamma_1}{\gamma_1-1} \mathcal{K}_1 T_1=
\frac{1}{2}v_2^2 + \frac{\gamma_2}{\gamma_2-1} \mathcal{K}_2 T_2 + Q
\end{equation}
where $\gamma$ is the ratio of specific heats, $Q$ is the energy
lost per unit mass during the shock process, and the equation of
state is applied as $p=(k/\mu m_H)\rho T = \mathcal{K}\rho T$. The
jump conditions (\ref{rh1})-(\ref{rh3}) enable one to solve the
J-shock structure.

We would be interested to consider the collision of two gas sheets
with velocities $v_0$ in the rest frame of laboratory. In this
reference frame, the post-shock will be at rest and the pre-shock
velocity is given by $v_1=v_0+v_2$, where $v_2$ is the shock front
velocity. Defining the pre-shock sound speed as $c\equiv
\sqrt{\mathcal{K}_1 T_1}$ and Mach number as $M_0 \equiv v_0/c$, the
equations (\ref{rh1})-(\ref{rh3}) lead to
\begin{equation}\label{temp1}
    \mathcal{T} = \frac{1}{\mathcal{R}} + \frac{M_0^2}{\mathcal{R}-1},
\end{equation}
\begin{equation}\label{temp2}
    \mathcal{T} = \frac{\gamma_1}{\gamma_2} \frac{\gamma_2 -1}{\gamma_1 - 1} +
    \frac{\gamma_2 -1}{2\gamma_2} \frac{\mathcal{R}+1}{\mathcal{R}-1}M_0^2 -
    \frac{(\gamma_2 -1)Q}{\gamma_2 c^2},
\end{equation}
where $\mathcal{T} \equiv \mathcal{K}_2 T_2 / \mathcal{K}_1 T_1$ and
$\mathcal{R} \equiv \rho_2 / \rho_1$ are the relative temperature
and density of the post-shock gas, respectively. Eliminating
$\mathcal{T}$ between the equations (\ref{temp1}) and (\ref{temp2}),
gives the square equation
\begin{equation}\label{cubicR}
    \mathcal{R}^2 +A_1 \mathcal{R} +A_2 =0,
\end{equation}
with coefficients as follows
\begin{eqnarray}\label{coeff}
  \nonumber A_1 &=& -\frac{\frac{\gamma_2+1}{\gamma_2-1}M_0^2
  + \frac{2\gamma_1}{\gamma_1-1}+ \frac{2\gamma_2}{\gamma_2-1}- \frac{2Q}{c^2}}{M_0^2+ \frac{2\gamma_1}{\gamma_1-1}
   - \frac{2Q}{c^2}}\\
  \nonumber A_2 &=& \frac{\frac{2\gamma_2}{\gamma_2-1}}{M_0^2+ \frac{2\gamma_1}{\gamma_1-1}
   - \frac{2Q}{c^2}}.
\end{eqnarray}

In the general case, the solution of equation (\ref{cubicR}) can
directly be found according to four parameters $\gamma_1$,
$\gamma_2$, $Q$ and $M_0$; then the relative temperature
$\mathcal{T}$ is obtained via equations (\ref{temp1}) or
(\ref{temp2}). The adiabatic shock is a special case with $Q=0$,
which in the strong supersonic collision ($M_0\rightarrow \infty$),
the equation (\ref{cubicR}) leads to $\mathcal{R}=4$ with assumption
of $\gamma_2=5/3$. In this case, the relative temperature is
limitless as $\mathcal{T} \approx M_0^2/3$. In fact, this
temperature is the maximum allowed value in a shock process, which
is $T_2^{max} = 3.86 \mu_2 \times 10^5 v_{07}^2 \mathrm{K}$ where
$v_{07} \equiv v_0/100 \mathrm{km.s^{-1}}$ (e.g., Hollenbach \&
McKee~1979). The most important parameter in the strong supersonic
shocks ($M_0>>1$) is the energy lost per unit mass during the shock
process, $Q=(n_2 \Lambda/\mu_2 m_H) t_{dur}$, where $\Lambda$
($\mathrm{erg.cm^{3}.s^{-1}}$) is the cooling function at the
post-shock region with density $n_2$, and $t_{dur}$ is the duration
time of the post-shock gas. Accurate determination of the cooling
time-scale requires specifying the elemental abundance of the
post-shock region, but a simple estimate can be obtained using
$t_{cool}\approx kT_2/(n_2\Lambda)$. Eliminating the $n_2\Lambda$,
we approximately have $Q/c^2\approx T (t_{dur}/t_{cool})$. If the
post-shock gas cools rapidly (i.e., $t_{cool}<<t_{dur}$), its
temperature cannot be grater than the molecular dissociation energy,
while in slow cooling rate (i.e., $t_{cool}>>t_{dur}$), the
temperature may even be grater than $10^4 \mathrm{K}$ causing the
atomic ionization process.

\begin{figure}
\begin{center}
\includegraphics[width=80mm]{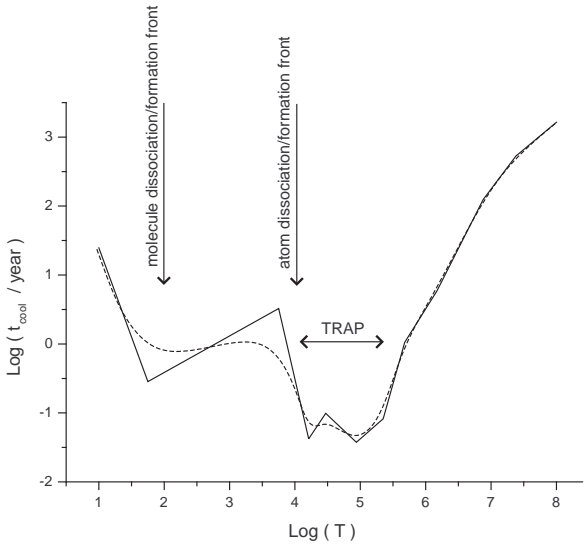}
\caption{The cooling time-scale as expressed in a piecewise linear
form for ionization degree $x_i=0.1$ and metallicities corresponding
to the solar neighborhood. The dotted curve is the B-spline fit to
this piecewise linear expression. The temperatures in which the
molecules and atoms dissociate/form are depicted as arrows, and the
minimum value of cooling time-scale is nominated as trap.}
\label{coolt}
\end{center}
\end{figure}

The cooling mechanisms take into account many different processes
that dominate in different ranges of temperature. We apply the
cooling function as outlined in the Figure~1 of Heitsch, Hartmann \&
Burkert~(2008) in which they used a combination of the rates quoted
by Dalgarno and McCray~(1972) and Wolfire et al.~(1995) for $T <
10^4 \mathrm{K}$, and the tabulated curves of Sutherland and
Dopita~(1993) for $T > 10^4 \mathrm{K}$. We can express the
logarithm of cooling function as a piecewise linear function of the
temperature logarithm as follows
\begin{equation}\label{cool}
    \Lambda = \Lambda_0 \left( \frac{T}{T_0} \right)^\beta,
\end{equation}
where $\Lambda_0$, $T_0$ and $\beta$ are given in Table~1 for
ionization degree $x_i=0.1$ and metallicities corresponding to the
solar neighborhood. In this way, the cooling time-scale can be
expressed in a piecewise form as
\begin{equation}\label{cooltime}
    t_{cool}\approx \frac{kT_0^\beta}{n_2 \Lambda_0} T^{1-\beta},
\end{equation}
which is shown in Fig.~\ref{coolt} for $n_2 \sim 10^4
\mathrm{cm^{-3}}$.


\begin{table}
\begin{center}
\caption{Parameters for the piecewise linear expression of the
cooling function, $\Lambda=\Lambda_0(T/T_0)^\beta$, with ionization
degree $x_i=0.1$ and with metallicities corresponding to the solar
neighborhood.\label{tbl}}
\begin{tabular}{|c|c|c|c|}
\hline\hline $\log T$ & $\Lambda_0 (\mathrm{erg.cm^3.s^{-1}})$ &
$T_0 (\mathrm{K})$ &
$\beta$ \\
\hline
$1.00 \rightarrow 1.75$ &  $1.74\times10^{-28}$  &  $10$    &    $3.59$\\
$1.75 \rightarrow 3.75$ &  $8.51\times10^{-26}$  &  $56$    &    $0.47$\\
$3.75 \rightarrow 4.21$ &  $7.41\times10^{-25}$  &  $5623$ &    $5.11$\\
$4.21 \rightarrow 4.47$ &  $1.66\times10^{-22}$  &  $16218$ &   $-0.42$\\
$4.47 \rightarrow 4.94$ &  $1.29\times10^{-22}$  &  $29512$ &    $1.89$\\
$4.94 \rightarrow 5.35$ &  $1.00\times10^{-21}$  &  $87096$ &    $0.17$ \\
$5.35 \rightarrow 5.68$ &  $1.17\times10^{-21}$  &  $223872$ &   $-2.36$\\
$5.68 \rightarrow 6.17$ &  $1.95\times10^{-22}$  &  $478630$ &   $-0.51$ \\
$6.17 \rightarrow 6.88$ &  $1.10\times10^{-22}$  &  $1479109$ &   $-0.87$\\
$6.88 \rightarrow 7.38$ &  $2.63\times10^{-23}$  &  $7585778$ &   $-0.26$\\
$7.38 \rightarrow 8.00$ &  $1.95\times10^{-23}$  &  $23988340$ &    $0.21$\\
\hline
\end{tabular}
\end{center}
\end{table}

When the strong shocking molecular gases proceed, at the first time,
the temperature increases near $T_2^{max}$ so that the molecules
will be dissociated and the atoms may be ionized. In the ionized
region, there is a trap in which $t_{cool}$ reaches to its minimum
value so that the plasma gas cools rapidly to recombine the
electrons and nucleons. Recombination becomes important when the
post-shock plasma cools to $10^4 \mathrm{K}$; a typical $H$ atomic
recombination time-scale is approximately $t_{rec} \approx
1/n_e\alpha_{rec}\approx10^5 /n_e (\mathrm{cm^{-3}}) \mathrm{yr}$,
where $\alpha_{rec}$ is the total recombination coefficient and
$n_e$ is the number density of electron (Osterbrock \& Ferland
2006). If the process of cooling mode continues, the atomic
post-shock gas will approach to the molecular gas. The $H_2$
molecule cannot form in the gas phase because it is not lonely able
to radiate away the excess energy of formation. The grains, which
have been survived in the shock process, can thus transfer this
extra energy of $H_2$ formation to their phonons (Hollenbach \&
Salpeter~1971). Although, the details of this mechanism are rather
complex, a simple estimation of $H_2$ formation time-scale is
$t_{mol} \approx 1/n_H k_m \approx 10^9/n_H^3 (\mathrm{cm^{-3}})
\mathrm{yr}$, where $k_m \approx 3\times 10^{-17}n_H(n_H+2n_{H_2})$
$\mathrm{cm^3.s^{-1}}$ is the hydrogen molecule formation rate on
the grain surfaces and $n_H$ is the hydrogen atomic number density
(Lequeux~2005). For a typical J-shock structure of a protostellar
accretion shocks, with $n_e \approx n_H \approx 10^4
\mathrm{cm^{-3}}$, we have $t_{rec} \approx 10 \mathrm{yr}$ and
$t_{mol} \approx 10^{-3} \mathrm{yr}$, thus the recombination occurs
slowly while the molecule formation occurs very fast. Although, we
find some insights for J-shock structure and its relaxation process,
but the shock process is rather a complex non-equilibrium thermal
and chemical mechanism. Thus, in the next section, we use the SPH
simulation to study the time evolution of processes in the strong
shocking molecular gases.


\section{Investigation by SPH}

We consider the head-on collisions of two molecular gas sheets. An
initial negative velocity is given to the particles with a positive
$x$-coordinate, and a velocity in the opposite direction to those
with a negative $x$-coordinate. Compositions of the sheets are
assumed as global neutral which consists of a mixture of atomic and
molecular hydrogen ($X\approx0.75$), helium ($Y\approx0.25$), and
traces of CO and other rare molecules. The mean molecular weight is
initially $1/\mu = X/2 + Y/4 \approx 0.4375$ for molecular case
($T<10^2 \mathrm{K}$), and in the post-shock region for simplicity
is assumed to be $1/\mu = X + Y/4 \approx 0.8125$ for atomic case
($10^2 \mathrm{K}< T < 10^4 \mathrm{K}$), and $1/\mu = 2X + 3Y/4
\approx 1.6875$ for ionized case ($T> 10^4\mathrm{K}$). The same
procedure is followed for the ratio of specific heats: $\gamma=7/5$
for diatomic molecular gas and $\gamma=5/3$ for atomic and ionized
cases. The chosen physical scales for length and time are $[l]=3.0
\times 10^{14} \mathrm{cm}=20 \mathrm{AU}$ and $[t]=3.0 \times
10^{7} \mathrm{s} = 1 \mathrm{yr}$, respectively, so the velocity
unit is approximately $100~\mathrm{km.s^{-1}}$. The gravitational
constant is set $G= 10^{-12} [l]^{3}.[t]^{-2}.[m]^{-1}$ for which
the calculated mass unit is $[m]=4.5 \times 10^{23} \mathrm{g}$.
Consequently, the physical scales for density and energy are
$[\rho]=1.6\times 10^{-20} \mathrm{g.cm^{-3}}= 10^{4} \mathrm{H.
cm^{-3}}$ and $[e]=4.5\times 10^{30}\mathrm{erg}$, respectively. Two
equal one dimensional molecular sheets with extension $x= 0.1 [l]$
is considered, which have initial uniform density and temperature of
$10^4 \mathrm{cm^{-3}}$ and $\sim 10 \mathrm{K}$, respectively.

The SPH method is well suited to address unbound astrophysical
problems, especially the behavior of gases subjected to compression
(Rosswog~2009). A worthy review of the SPH methodology and its
applications can be found in Monaghan~(2005). In this method, fluid
is represented by $N$ discrete but extended/smoothed particles (i.e.
Lagrangian sample points). The particles are overlapping, so that
all the involved physical quantities can be treated as continuous
functions both in space and time. Overlapping is represented by the
kernel function, $W_{ab} \equiv
W(\textbf{r}_a-\textbf{r}_b,h_{ab})$, where $h_{ab} \equiv
(h_a+h_b)/2$ is the mean smoothing length of two particles $a$ and
$b$. The density is estimated via usual summation over neighboring
particles,
\begin{equation}\label{sphden}
\rho_a=\sum_b m_b W_{ab},
\end{equation}
the acceleration equation in the one-dimensional usual symmetric
form is
\begin{equation}\label{sphacc}
\frac{d v_a}{dt}=-\sum_b m_b (\frac{p_a}{\rho_a^2}+
\frac{p_b}{\rho_b^2}+ \Pi_{ab})  \frac{dW_{ab}}{dx_a},
\end{equation}
and the SPH equivalent of the energy equation is
\begin{eqnarray}\label{sphenergy}
\nonumber \frac{du_a}{d t}=\frac{1}{2} \sum_b m_b
(\frac{p_a}{\rho_a^2} +\frac{p_b}{\rho_b^2} +\Pi_{ab}) v_{ab}\frac{d
W_{ab}}{d x_a}\\ -\frac{\Lambda_0}{(\mu_a m_H)^2} \rho_a \left(
\frac{T_a}{T_0}\right)^\beta,
\end{eqnarray}
where $u_a=\frac{1}{\gamma_a-1}\mathcal{K}_aT_a$ is the thermal
energy per unit mass, and $\Pi_{ab}$ is the artificial viscosity
between particles $a$ and $b$
\begin{equation}\label{av}
\Pi_{ab}=\cases{
       \frac{-\alpha^* v_{sig} \mu_{ab}^*
       +\beta^* \mu_{ab}^{*2}}{\bar{\rho}_{ab}}, &
       if $\textbf{v}_{ab}.\textbf{r}_{ab}<0$,\cr
       0 , & otherwise,}
\end{equation}
where $\textbf{v}_{ab}\equiv \textbf{v}_a - \textbf{v}_b$ and
$\textbf{r}_{ab}\equiv \textbf{r}_a - \textbf{r}_b$ are relative
velocity and the distance of particles, $\bar{\rho}_{ab}=
\frac{1}{2}(\rho_a+\rho_b)$ is an average density, $v_{sig} =
(c_a+c_b)/2$ is signal velocity where $c_a$ and $c_b$ are the sound
speed of particles, and $\mu_{ab}^*$ is defined as its usual form
\begin{equation}\label{muab}
\mu_{ab}^*=\frac{\textbf{v}_{ab} \cdot\textbf{r}_{ab}}{h_{ab}}
\frac{1}{r_{ab}^2/h_{ab}^2+\eta^2}
\end{equation}
with $\eta\sim 0.1$.

\begin{figure}
\begin{center}
\includegraphics[width=80mm]{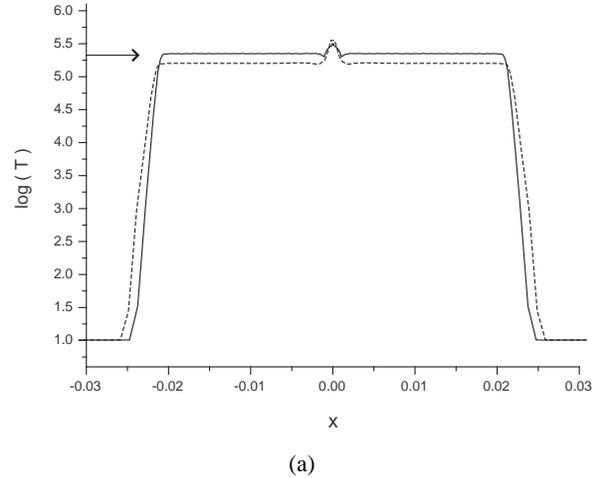}\\(a)\\
\includegraphics[width=80mm]{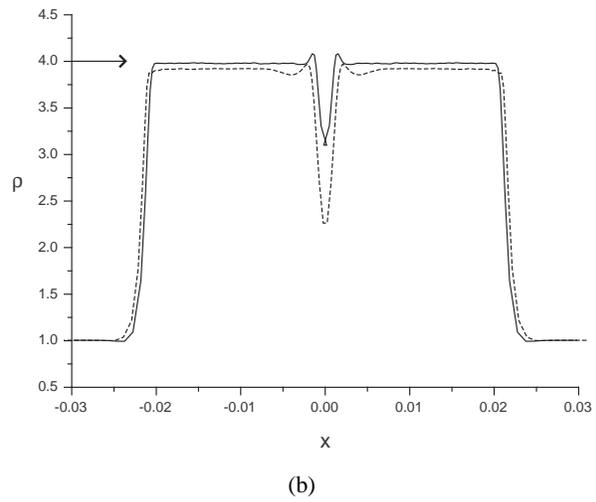}\\(b)\\
\caption{(a) Temperature and (b) density of the adiabatic shock
($\Lambda_0 = 0$) in the head-on collision of two sheets with
initial Mach number $M_0=500$. The solid curves are derived with
variable viscosity (\ref{alpha}) and (\ref{beta}), while the dotted
ones are from the common artificial viscosity of Monaghan~(1989)
with $\alpha^*=1$ and $\beta^*=2$. The analytical results of strong
supersonic adiabatic shocks are depicted by
arrows.}\label{tempdenpos}
\end{center}
\end{figure}

To prevent unphysical solutions with inter-particle penetration and
unwanted heating, we use $\alpha^*$ and $\beta^*$ in the form of
variables with respect to time,
\begin{equation}\label{alpha}
\frac{d \alpha^*_a}{dt} = - \frac{\alpha^*_a -
\alpha_{\mathrm{min}}} {\tau_a} + z_\alpha \mathcal{S}_a,
\end{equation}
and
\begin{equation}\label{beta}
\frac{d \beta^*_a}{dt} = - \frac{\beta^*_a - \beta_{\mathrm{min}}}
{\tau_a} + z_\beta \mathcal{S}_a,
\end{equation}
respectively, where $\mathcal{S}_a=\max (- \frac{dv_a}{dx_a},
\frac{\dot{\rho}_a}{\rho_a} - \frac{\dot{h}_a}{h_a} -
\frac{v_{\mathrm{sig}}}{r_{ab}}, 0)$ is the restricted source term,
$\tau_a \equiv h_a / (\mathcal{C} v_{sig})$ with $0.1 < \mathcal{C}
< 0.2$ is the decay time-scale, and the parameters $z_\alpha$ and
$z_\beta$ are chosen to regulate the effect of source term so that
the heat production and post-shock oscillations are controlled in
the numerical simulations (Nejad-Asghar, Khesali \& Soltani~2008).
Here, we choose $\alpha_{min}=1$, $\beta_{min}=2$,
$\mathcal{C}=0.2$, $z_\alpha=3$, and $z_\beta=0.01$ for the best
smoothed results of simulations (see, e.g., Fig.~\ref{tempdenpos}).
Clearly, comparison between numerical and analytical results in
Fig.~\ref{tempdenpos} verifies the accuracy and physical consistency
of the employed numerical tool.

\begin{figure}
\includegraphics[width=80mm]{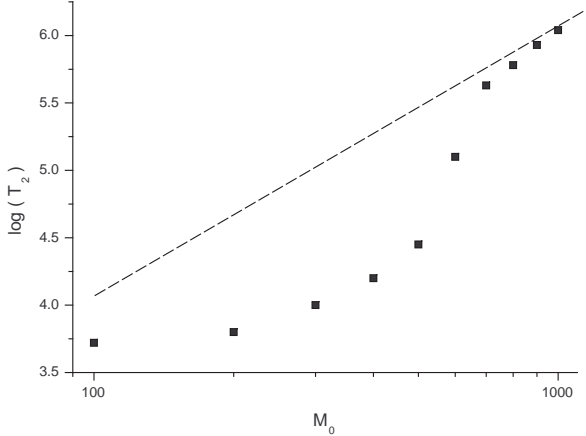} \caption{Post-shock
temperature at the simulation epoch for different initial Mach
numbers. The adiabatic case, $T_2 \propto M_0^2$, is shown by
dash-line.}\label{tempmach}
\end{figure}

The width of the post-shock region increases by time, until it
reaches to all extensions of the simulated sheets, or the
temperature of center of the shocked region cools less than
$10\mathrm{K}$. For computer experiments which are considered here,
with initial extension $x = 0.1 [l]$, we stop the simulation when
$90\%$ of total SPH particles enter to the post-shock region or the
temperature of center of the shocked region cools less than
$10\mathrm{K}$. In this simulation epoch, the cooling rate affects
on the post-shock temperature as shown in Fig.~\ref{tempmach} for
various Mach numbers. In this figure, the relationship $T_2 =
\frac{\mu_2}{\mu_1} \frac{T_1}{3} M_0^2$, as mentioned for the
strong supersonic adiabatic shocks, is depicted by dash-line. We see
the cooling rate causes to decrease the value of post-shock
temperature. The effect of cooling rate appears more clear in the
Mach numbers which cause to set the post-shock temperature in the
trap region as outlined by Fig.~\ref{tempdenpos}. Since for larger
Mach numbers, the simulation epoch (i.e, the time in which $90\%$ of
SPH particles of our simulation enter to the post-shock region or
the temperature of center of the shocked region cools less than
$10\mathrm{K}$) is shorter, the post-shock temperature reaches
asymptotically to the adiabatic case.

\begin{figure}
\includegraphics[width=80mm]{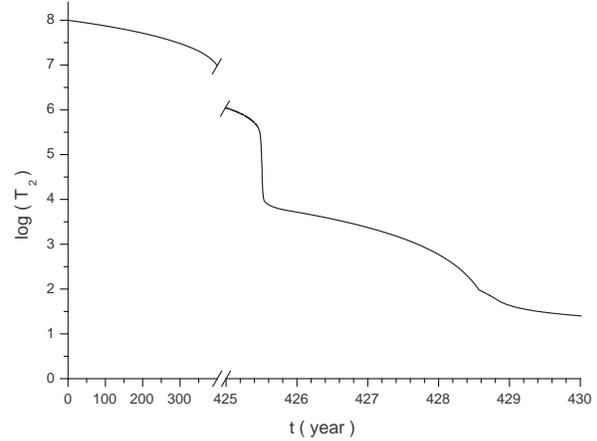} \caption{Relaxation of the
temperature of the post-shock gas in the head-on collision of two
sheets.}\label{temptime}
\end{figure}

In the general case, the high temperature of the post-shock region
lead to splash it into the medium. But, in the accretion shock, the
infall matter confides the shocked gas so that it may cool and end
up with an accretion rotating disk. For finding the relaxation time,
we use the equation (\ref{sphenergy}) with $v_{ab}\approx 0$, thus,
we have
\begin{equation}\label{temprelax}
\frac{1}{\gamma_a-1}\mathcal{K}_a\frac{dT_a}{d t}=
-\frac{\Lambda_0}{(\mu_a m_H)^2} \rho_a \left(
\frac{T_a}{T_0}\right)^\beta.
\end{equation}
The temperature of the accreted shocked gas can be obtained by
integrating the equation (\ref{temprelax}). The result is in a
piecewise change from $T_{a1}$ (at $t_1$) to $T_{a2}$ (at $t_2$) as
follows:
\begin{eqnarray}\label{tempchange}
  \nonumber T_{a2}=\{ T_{a1}^{1-\beta}- \frac{(1-\beta)(\gamma_a-1)}{\mathcal{K}_a}
  \frac{\Lambda_0}{(\mu_a m_H)^2}\times \\ \frac{\rho_a}{T_0^\beta} (t_2-t_1)
  \}^\frac{1}{1-\beta},
\end{eqnarray}
which is shown in Fig.~\ref{temptime}, with assumption of
$\rho_a\approx 4$.

\section{Summary and conclusions}
Molecular cloud cores are rotating so that in the process of
collapse and protostellar formation, the infalling matters, which
arrive at the equator, collide and dissipate the kinetic energy of
motion perpendicular to the equatorial plane so that an accretion
disk can be formed. The speed of infalling matters at the equator is
so high that the strong supersonic shocks appear, and the
temperature of post-shock gas is increased so causing to dissociate
the molecules and ionize the atoms. In adiabatic strong supersonic
shocks, the density of post-shock region is about fourfold of
initial density, and the temperature is increased proportional to
the square of the Mach number. On the other hand, the cooling
processes of the post-shock gas can decrease the temperature so that
the ionized gases can be recombined to form the atoms and molecules,
if the duration time-scale of the post-shock region is longer than
the cooling time-scale. The suitable cooling function for the
post-shock gas (Table~1) showed that the cooling time-scale has a
minimum which is at the ranges of ionizing temperature
(Fig.~\ref{coolt}). This minimum, which is nominated as a trap,
causes to cool the post-shock gas in a fast rate. Thus, if the
initial speed of colliding matters is so high that the temperature
of post-shock gas settles in this trap region, it will quickly cool
and electrons recombine with nucleons.

For investigating the time evolution of the post-shock gas in the
strong supersonic collisions, which occurs at protostellar accretion
shocks, we used the SPH method with the variable artificial
viscosity to obtain the more smoothed results of simulations
(Fig.~\ref{tempdenpos}). The simulations of strong shocks with great
Mach numbers show that the temperature of the post-shock gas is
quickly increased near to $T_2^{max}$, which is for the adiabatic
case, and gradually decreased by the time according to the cooling
rate. Decrease of temperature of the post-shock gas at the
simulation epoch is shown in Fig.\ref{tempmach} for different values
of the Mach numbers. We see from this figure that the decreasing
rate of temperature is very fast in the trap region as depicted by
Fig.~\ref{coolt}.

The simulations show that the temperature of center of the
post-shock gradually decreases, while in the ridges, it stays about
$T_2^{max}$ because of continuously infall of matters. Temperature
decrease of the central region leads to increase of its density in
an isobaric manner so that an accretion thin disk can be formed.
Thus, over the time, center of the collisional infalling matters
(i.e., equatorial plane) converts to a dense molecular thin disk
with an atomic envelope and ionized gas which comes from strong
shocks of continuous infalling matters. The time in which this
structure occurs, depends on the Mach number that is shown in
Fig.~\ref{temptime}. The temperature range of the trap, which leads
to fast cooling of the post-shock gas, is clearly seen in
Fig.~\ref{temptime}, too. Thus, we see that the cooling processes of
the post-shock gas in the protostellar accretion shocks can lead to
the formation of an accretion thin molecular disk at the equatorial
plane, in a convenient time-scale. This dense molecular thin disk is
appropriate for grain coagulation and formation of proto-planetary
entities.

\section*{Acknowledgments}
This work has been supported by grant of Research and Technology
Deputy of University of Mazandaran.

\end{document}